\documentclass[aps,prl,showpacs,twocolumn,nofootinbib]{revtex4-1}
\usepackage[dvips]{graphicx}
\begin{document}
\title{Comment on Perez et al PRX 2, 041005 (2012)}
\author{A. Beresnyak}
\affiliation{Los Alamos National Laboratory, Los Alamos, NM, 87545}
\affiliation{Ruhr-Universit\"at Bochum, 44780 Bochum, Germany.}

\begin{abstract}
Recently Perez et al \cite{Perez2012} wrote on the spectral slope of MHD turbulence claiming that it is consistent with -3/2.
This work contains a number of errors, factual inaccuracies and puzzling methods in the interpretation
of numerics.  We argue that the numerical evidence that the authors presented is
actually against -3/2 slope and the assumption of universal alignment, opposite to what they claim. Perhaps the most puzzling is the
Fig.~8 that claims that the authors measured the inertial range length and that it is consistent with $Re^{2/3}$
scaling. At a close inspection it appears that the datapoints are not the result of a measurement, but rather
were calculated by a formula.
\end{abstract}

%\begin{keywords}
%MHD -- turbulence.
%\end{keywords}
%\keywords{MHD -- turbulence}
\maketitle

Currently there were conflicting reports regarding the spectral slope of MHD turbulence. While Beresnyak reports numerical data as roughly
consistent with $-5/3$ slope and inconsistent with $-3/2$ \citep{B11, B12b}, Perez et al \cite{Perez2012} group claims it is consistent with $-3/2$.
Both groups do driven simulations of strong turbulence in the strong mean field limit by using pseudospectral code. For dissipation scheme
Beresnyak used hyperdiffusion as well as normal diffusion, while Perez et al used only normal diffusion. 
%While Beresnyak claims that his data is compatible
%with what was reported earlier by Perez et al, the latter group claims otherwise and claim that Beresnyak's result is dominated by numerical error.

Recently Perez et al published a detailed paper on their latest high-resolution simulations with long evolution in time, and also claimed
that Beresnyak's result is dominated by numerical error. The paper is the subject of this comment.

In this comment we discuss several major issues with Perez et al interpretation of numerics, including:
a) the issue of Fig.~8, b) the claim of the universality of alignment, c) the claim of $-3/2$ slope, d) the claim of the "fake" numerical convergence,

{\it The issue of Fig.~8} --- On Fig.~8 the Authors claimed that they measured the length of the inertial range and that its scaling with Reynolds
number is consistent with Boldyrev's model \cite{boldyrev2005,boldyrev2006}. On close inspection of Fig.~8 it is evident that the Authors arbitrarily took a constant $k_\perp=4$ as the
beginning of the inertial range and the constant $k_\perp \eta =0.1$ as the end of it. The length of the inertial range that the Authors ``measured''
and presented in the lower panel of Fig.~8 was, therefore, calculated by a formula $l_0/l_d=40 \eta^{-1}$, where $\eta$ is defined by Eq.~8,
so $l_0/l_d=40 \epsilon^{2/9} \Lambda^{-1/9} \nu^{-2/3}$. All quantities that enter this formula are the parameters of the simulations, so this ``measurement``
does not constitute a real measurement. Also with $\epsilon^{2/9}$ vary only by a factor of 0.99 and $\Lambda^{1/9}$ vary by a factor of 0.99,
the main dependence of $l_0/l_d$ is, automatically, $\nu^{-2/3} \sim Re^{2/3}$, which was confirmed on the lower panel and incorrectly attributed
to the correspondence with Boldyrev's model.
If one does {\it a real measurement} assuming, e.g., that the inertial range length is the greatest stretch of the spectra, deviating
from $-3/2$ law by no more than 5\%, then the scaling of such inertial range length is grossly inconsistent with $Re^{2/3}$, see Fig~1 of this comment.
The same is true if some other criterion is used, e.g., using deviation of 10\%.

We conclude that the Author's argumentation that numerics exhibit $Re^{2/3}$ scaling of the inertial range length is errorneous and void.
Instead, the measurements of the inertial range length show a non-power law dependence on $Re$. 

As a side-note, the length of the $-3/2$ range that Perez et al simulations seem to saturate to with increasing Re, $8\div10$, is consistent
with the length of the $-3/2$ range published in \citep{B11, B12b}.

\begin{figure}
\includegraphics[width=0.85\columnwidth]{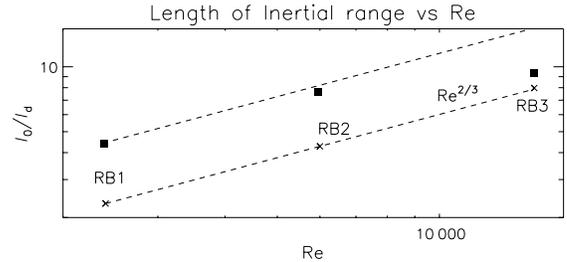}
\caption{The inertial range length reported by Perez et al (crosses), which they calculated by a formula and the overplotted
results of the real measurement of such length (squares),
assuming that the inertial range length is the greatest stretch of the spectra, deviating
from $-3/2$ law by no more than 5\%.}
\label{f1}
\end{figure}

{\it The claim of the -3/2 slope} --- Perez et al simulations feature rather long evolution in time, 100 dynamical times on the outer scale.
These simulations, therefore, will produce measurements with significantly lower
statistical error than that of others, which typically used 10-20 outer dynamical times\footnote{It have to be noted that smaller scales have much
better statistics because they have smaller timescales and much more realizations in the simulation box, which is
why 10-20 outer dynamical times is usually considered adequate for the turbulence scaling studies.}.
For the purpose of this comment we will only assume that the statistical error in Perez et al is lower than that in previous hydrodynamic
or MHD simulations, such as \cite{yeung1997, gotoh2002, kaneda2003, B11, B12b}.
The scaling convergence argument which has been used in the above studies require that all curves from different simulations collapse
into a single curve starting with some scale $l$ and {\it on all smaller scales}, including dissipation scales. If there is a convergence in the inertial
range but there is no convergence on the viscous scales, this would mean that there is some source of systematic error on grid scale that
actually depends on the resolution of the experiment, such as the numerical equations on grid scale are actually different for
$1024^3$ and $2048^3$. The author's Fig~8 does show such a lack of convergence within $k_\perp \eta =0.08\div 1.0 $.
The authors, however, neither mention their non-convergence on these scales, 
nor they claim any such strange systematic error. This error has never been reported before and \cite{yeung1997, gotoh2002, kaneda2003, B11, B12b}
always report that the curves collapse onto the same curve starting with some scale $l$ and on all smaller scales.
To the best of our knowledge such systematic error simply does not exist and true convergence should always be in the above conventional sense.

As a curious excersize one can also check that if the perpendicular spectrum converges on the inertial range but does not converge
on viscous scales, the other type of spectra, e.g. one-dimensional spectrum will not converge at all\footnote{One-dimensional spectrum is
defined as a power spectrum of a quantity sampled along a certain line and then averaged over all lines and directions. This type
of spectrum is of interest because it is produced by the solar wind measurements from satellites. The mathematical relation between
isotropic or perpendicular spectrum $E(k)$ and the one dimensional spectrum is $P(k)=\int^\infty_k E(k_1) \frac{dk_1}{k_1}$ \cite{monin1975}.
From this expression it is clear that if two $E_1(k)$ and $E_2(k)$ are coincident in the inertial range but different on viscous scale, the corresponding $P_1(k)$
and $P_2(k)$ are different everywhere. Note that \cite{boldyrev2011} confusingly compare simulated E(k) and the measured P(k).}.
So, in this case the fact of the convergence
hangs on the type of spectrum one uses, which is highly doubtful. In other words, Fig~8. does not demonstrate convergence of spectra with -3/2
scaling, on the contrary, it demonstrates the lack of such convergence. 

Also, it is quite puzzling that while in sections 2-3 the authors spent a great deal of time explaining why simulations with reduced
parallel resolution are bad, they use the very same reduced parallel resolution simulations for convergence study, even though they have
data from full parallel resolution simulations? Another puzzling feature of Fig.~8 is the high level of fluctuations of RB3 spectrum.
All other spectra, presented in this paper are very smooth. The authors could have used a smaller time evolution
for this spectrum but they never say so in the paper.

{\it The claim of the universality of alignment} --- Similar convergence arguments can be used to claim a universal scaling of some other
quantity. The authors were trying to support universality of so-called dynamic alignment $\theta(l)$. Again their figure Fig.~7 does not show convergence and
having very low level of statistical error and no source of systematic error the authors should have been expecting rescaled dynamic alignment curves to collapse
onto a single curve. They, however, do not, quite consistently with \cite{B12b}. Contrary to the numerical evidence the authors still claim that $\theta(l)$ scalings
are universal. 

{\it The logical loop in the alignment universality claim} -- The Authors bring several qualitative arguments in favor of universality
of their alignment measure $\theta(l)$, however, before they do a scaling study
to demonstrate such universality, they say that they rather postulate it and redefine outer scale $\Lambda$ so that $\theta(l)$ will fit better between
simulations with different resolutions and after that say that alignment is ``remarkably stable''. This argumentation is, obviosly, a logical loop. 
But even after the Authors fitted their $\Lambda$, the $\theta(l)(\Lambda/\eta)^{1/4}$ still didn't converge as noted above (Fig.~7).

Defining outer scale by alignment is certainly unusual.
Conventionally defined outer scale, $L=(3\pi/4E)\int_0^\infty k^{-1} E(k) dk$ is a robust measure that shows only minuscule dependence on resolution,
as long as the driving procedure is unchanged. The Author's  $\Lambda$, defined by alignment, change considerably, however. The Authors did not comment
of whether they explicitly changed driving procedure between three simulations RB1a-3a or why $\Lambda$ could be different in each of these simulations,
especially considering that $\epsilon$ is virtually constant.

Obviously, the $\Lambda$, defined by the Authors, depends on resolution, because alignment {\it is not} universal.

{\it The claim of the "fake" numerical convergence} --- In their Appendix the Authors claim that the convergence on viscous scales observed in \cite{B11, B12b}
is a ``fake'' or a numerical artefact convergence which is due to numerical error. This is fairly puzzling claim, considering the scaling study is a traditional method
which has been used for years, e.g. \cite{yeung1997, gotoh2002, kaneda2003}. Also it is quite puzzling because the convergence on viscous scales from
different simulations when one compensates spectrum by Kolmogorov scaling and plots it against $k\eta_{41}$ simply means that the velocity perturbation
on viscous scale is proportional to the Kolmogorov velocity\footnote{Indeed, if the spectrum $E(k)$ is compensated by $\epsilon^{-2/3} k^{5/3}$ and converge
to some constant $C$ at $k \eta_{41}$ equal to, say, 0.2, then the velocity perturbation $\delta v = \sqrt{E(k)k}$ on this scale will be determined by
$C^{1/2}0.2^{-1/3} (\nu \epsilon)^{1/4}$.}
$v_{\eta 41}=(\nu \epsilon)^{1/4}$. Why would numerical error conspire in such a way as to produce such specific dependence
of perturbation amplitude on $\epsilon$ and $\nu$? Why such a puzzling phenomenon has never been reported before?

In fact the scaling study is relatively unaffected by what happens on small scales, because it is just a rescaling argument\cite{B12b}. The difference
between rescaled quantities on the viscous scale have very little to do with what happens on viscous scales, but has to do with the difference in resolutions and
the suggested rescaling slope. For example, taking $1024^3$ and $2048^3$ simulations the latter spectrum has to be shifted left by a factor of 2 and the spectrum
has to be shifted up by a factor of $2^{5/3}$. Why would numerical error know that it has to produce a factor of $2^{5/3}$ less numerical
noise on the grid scale in $2048^3$ compared to $1024^3$? Why would numerical error on the grid scale be so different in these two simulations and
why would it be different by this precise factor?

The Authors also claim that choosing $N \eta_{41} =const$ resolution criteria, i.e. the conventional criteria that people used before, will
result in some ``numerical convergence'', which ``should not be confused with the convergence to the physical solution''. It seems that the Authors are
unaware of the standard resolution studies that were performed to confirm the convergence of spectra by increasing resolution
and keeping all other parameters constant. In a proper scaling study each individual spectrum from each
simulation is converged to the physical solution within some error. Therefore, convergence of numerical spectra means absolute convergence.
Ironically, the Authors themselves
use $N \eta_{41} =const$ resolution criteria and never comment on why they did not believe their own model, which they claim
their data supported, and use $N \eta_{41} =const$ instead of $N \eta =const$.

The Authors also claimed that their simulations are better resolved than \cite{B11, B12b}. In doing so they referred to the simulations R8 and R9 in \cite{B12b}
which are the only simulations presented in that paper, which are under-resolved by conventional criteria. They do not mention other simulation groups,
such as R6-7, which are quite similar in numerical setup to what the Authors presented in their resolution study in Fig.~8, i.e. RB1a-RB3a.
Indeed R6 is $384\times 1024^2$, while RB2a is  $256\times 1024^2$, R7 is $768\cdot 2048^2$, while RB3 is $512\cdot 2048^2$. Moreover, R6-7 have
lower Reynolds numbers and R6-7 has $k_{\rm max} \eta_{41} = 0.95$, while in  RB1a-3a $k_{\rm max} \eta_{41} = 0.8$, i.e. R6-7 are
{\it better} resolved by a factor of 1.19, and, at the same time, have higher parallel resolution than RB1a-3a. In other words, when comparing
what the authors presented in their resolution study on Fig.~8 and the resolution study of R6-7 in \cite{B12b}, the latter are better resolved in both
parallel and perpendicular directions. The Authors, however, never mention this and made it sound like the simulations in \cite{B12b} are under-resolved
compared to theirs.

The Authors also make vague analogies between simulated truncated Euler equations that shows ``thermalized'' $k^2$ tail and the driven hyperviscous simulations.
At a close inspection these analogies do not hold. Indeed, the $k^2$ tail is essentially non-stationary, as the energy is being dumped from the turbulent
cascade into this tail. One can only speak of a certain snapshot of such simulation whose spectra might be similar to the spectrum of the very high
order hyperviscosity with gigantic botteneck bump. Out of simulations presented in \cite{B11, B12b} only one has a visible bottleneck bump and its
relative amplitude, 0.22 is even lower than the amplitude of bottleneck in hydrodynamics with normal viscosity, 0.31. Then, why hydrodynamic viscous
scales are not similar to the ``thermalized'' tail and our viscous scales are? Furthermore, the Authors speculate that the short $-5/3$ range
observed \cite{B11} is the result of an anti-bottleneck effect. But that would require that the anti-bottleneck effect has an amplitude comparable to the amplitude
of the bottleneck effect itself. This has never been observed and the anti-bottleneck effect that the authors call ``pseudodissipation``
is always much weaker than bottleneck effect, is not noticeable in viscous hydro with its 0.31 amplitude and is not supposed to be visible
in our simulation R3 with its 0.22 amplitude of the bump.
Needless to say, no scaling study argument is applicable to the ``thermalized'' tail, which is not even statistically
stationary and this tail is not actually thermalized \cite{ray2011}.
As to the numerically resolved statistically stationary simulations with hyperviscosity, the scaling study argument is still well-applicable \cite{B12b}.

Furthermore, the authors claim that the alignment is ``partially lost'' in the ``thermalized`` region. This is, again, puzzling, because the alignment is not
lost, i.e. its value does not equal to the value corresponding to the random vectors of ${\bf v, B}$, but it just flattens out, i.e. becomes
independent on scale as was clearly presented in \cite{B11}.

Finally, on their last plot of the Appendix the Authors make a convergence study between simulations with different geometries of the grid cell, RB2c and RB3a.
They don't obtain any convergence with either -3/2 or -5/3. What can be derived from such a study is unclear, because the scaling study
argument \cite{gotoh2002,B11,B12b}
simply does not work in this case and one is not supposed to do convergence study between such simulations. The Authors, nevertheless, claim that this
lack of convergence somehow supports their argumentation.

\begin{figure}
\includegraphics[width=1.0\columnwidth]{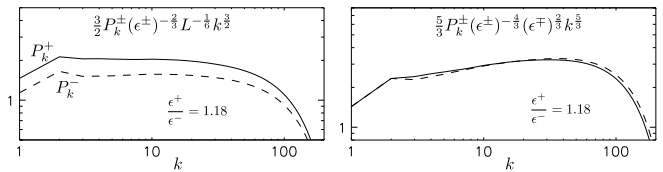}
\caption{The spectra for imbalanced simulation reported in \cite{BL10}, compensated by the factors corresponding to \cite{PB09} model and
\cite{LGS07} model.}
\label{f2}
\end{figure}

{\it Imbalance spectra} -- The Authors refer to previous studies of imbalanced case as producing ``conflicting results'' and claim that their data
further support the -3/2 spectral slope. However, as we see previously, even the balanced slope is not agreed upon and neither of the Author's
spectra shows clear -3/2 slope. In this situation a rigorous resolution study is, again, necessary to confirm or reject models. 
In particular, one wants to compensate spectrum by $E(k) k^\pm(\epsilon^\pm)^{-2/3}L^{-1/6}k^{3/2}$ if \cite{PB09} is correct or
by $E(k)^\pm(\epsilon^\pm)^{-4/3}(\epsilon^\mp)^{2/3}k^{5/3}$ if \cite{LGS07} is correct. Fig.~2 of this comment shows such comparison from a 
low-imbalance simulation previously reported in \cite{BL10}. As we see the \cite{PB09} model is grossly inconsistent with numerics. Unfortunately,
the Authors neither mention this nor they discuss how the predictions of \cite{PB09} should be modified in order to be consistent with numerics.
Furthermore, citing ``conflicting results'' from earlier simulations in the Introduction the Authors significantly distort the literature by citing
these earlier simulations as ``strongly imbalanced''. The Authors themselves presented simulations with energy imbalance of around 3,
while \cite{BL10} presented data for energy imbalance down to 1.35.

{\it Alignment controversy} -- The Authors define their alignment by Eq.~4. Why this specific function has to be relevant for the spectral slope
is unclear. The Authors previously claimed in \cite{Boldyrev2009b} that this alignment measure reduces the energy transfer function 
$\langle\delta w^{\mp}_{l\|} (\delta w^{\pm}_{l})^2 \rangle$ from \cite{politano1998}, but this was just a plausibility argument equivalent
to saying that alignment essentially causes
anticorrelation of $w^{\pm}$ reducing the above function. Similar plausibility arguments can be applied to other alignment measures,
such as $IM=\langle |\delta (w^+_\lambda)^2- \delta (w^-_\lambda)^2|\rangle /\langle \delta
(w^+_\lambda)^2+ \delta (w^-_\lambda)^2\rangle$ reported in \cite{BL09b,B12b}. However, IM does not show such a high scale-depencency
as the Author's alignment. Why their alignment measure is so special, compared to other measures, apart from showing approximate
$l^{1/4}$ scaling is never discussed by the Authors.

\par\ \ \ 

\par

%tau_corr=0.97

\def\apj{{\rm ApJ}}           
\def\apjl{{\rm ApJ }}          
\def\apjs{{\rm ApJ }}          
\def\grl{{\rm GRL }}
\def\aap{{\rm A\&A } }
\def\mnras{{\rm MNRAS } }
\def\physrep{{\rm Phys. Rep. } }               
\def\prl{{\rm Phys. Rev. Lett.}} 
\def\pre{{\rm Phys. Rev. E}} 
\bibliography{all}

\end{document}